\documentclass[]{aa}

\usepackage{graphicx}             
\usepackage{float}                 
\usepackage{placeins}              
\usepackage{adjustbox}             
\usepackage{lscape}                
\usepackage{wrapfig}               
\usepackage{epstopdf}              
\usepackage{subfigure}             
\usepackage{booktabs}              
\usepackage{amsmath}               
\usepackage[T1]{fontenc}           
\usepackage{float}                 
\usepackage{gensymb}               
\usepackage{rotating}              
\usepackage{siunitx}
\usepackage{dcolumn}
\usepackage{hyperref}

\usepackage{txfonts}
\usepackage{natbib}
\usepackage{graphicx}  
\usepackage{amsmath}   
\usepackage{multirow}  

\bibpunct{(}{)}{;}{a}{}{,} 

\DeclareRobustCommand{\VAN}[3]{#2}
\let\VANthebibliography\thebibliography
\def\thebibliography{\DeclareRobustCommand{\VAN}[3]{##3}\VANthebibliography}

\DeclareGraphicsExtensions{.ps}
\usepackage{graphicx}
\usepackage{txfonts}
\begin{document}

\title{Cyclical accretion regime change in the slow X-ray pulsar 4U 0114$+$65 observed with \textit{Chandra}. }

\author{
Sanjurjo-Ferr\'{i}n, G.$^{1}$,
Torrej\'on, J.M.$^{1}$,
Postnov, K.$^{2}$,
Nowak, M.$^{3}$
Rodes-Roca, J.J.$^{1}$,
Oskinova, L.$^{4}$,
Planelles-Villalva,J.$^{1}$,
Schulz, N.$^{5}$,
}

\institute{
$^{1}$Instituto Universitario de F\'{i}sica Aplicada a las Ciencias y las Tecnolog\'ias, Universidad de Alicante, 03690 Alicante, Spain\\
$^{2}$ Sternberg Astronomical Institute, Moscow M.V. Lomonosov State University, Universitetskij pr, 13, Moscow 119234, Russia\\
$^{3}$ Department of Physics, Washington University in St. Louis, Missouri, USA\\
$^{4}$Institute for Physics and Astronomy, Universit\"{a}t Potsdam, 14476 Potsdam, Germany\\
$^{5}$ MIT Kavli Institute for Astrophysics and Space Research, Cambridge, Massachussetts, USA\\
}

\date{Received XXX / Accepted XXX}

\abstract
{4U 0114+65 is a high-mass X-ray binary system formed by the luminous supergiant B1Ia, known as V{*} V662 Cas, and one of the slowest rotating neutron stars (NS) with a spin period of about 2.6 hours. This fact provides a rare opportunity to study interesting details of the accretion within each individual pulse of the compact object.
In this paper, we analyze 200 ks of \textit{Chandra} grating data, divided into 9 uninterrupted observations around the orbit. The changes in the circumstellar absorption column through the orbit suggest an orbital inclination of $\sim$ $40^{\circ}$ with respect to the observer and a companion mass-loss rate of $\sim$ 8.6 10$^{-7}$ $M\sun$ yr$^{-1}$. The peaks of the NS pulse show a large pulse-to-pulse variability. Three of them show an evolution from a brighter regime to a weaker one. 
We propose that the efficiency of Compton cooling in this source fluctuates throughout an accumulation cycle. After significant depletion of matter within the magnetosphere, since the settling velocity is
$\sim \times$ 2 times lower than the free-fall velocity, the source gradually accumulates matter until the density exceeds a critical threshold. This increase in density triggers a transition to a more efficient Compton cooling regime, leading to a higher mass accretion rate and consequently to an increased brightness.}


\titlerunning{4U }

\authorrunning{Sanjurjo-Ferr\'{\i}n et al. }

\maketitle


\section{Introduction.}

High Mass X-ray binaries (HMXRBs) are systems formed by a compact object (i.e. a neutron star or a black hole) orbiting a massive star (the companion). These are excellent laboratories where the process of accretion and the structure of the companion's stellar wind, can be studied \citep{2017SSRv..tmp...13M}. The study HMXRBs is important to understand the evolution of close  binaries, which may become compact-object mergers and eventually sources of gravitational waves and/or  short $\gamma$-ray bursts. They also provide insight into the behavior of matter at extreme gravitational and magnetic fields. 
Understanding these processes is fundamental to modern astrophysics and has been the driver of multiple theoretical and observational studies. 

The HMXB 4U 0114$+$65 was discovered by \cite{1977IAUC.3144....2D} in the SAS 3 Galactic survey. The companion star is V$^{*}$ V662 Cas, a luminous BI1a supergiant \citep{Reig}. With an orbital period of $\sim 11.6$ d, the neutron star (NS) orbits the companion deeply embedded into its wind at an orbital radius of $\approx 1.34-1.65 R_{\star}$, thereby offering the opportunity to probe the inner wind of the B1 supergiant star. It shows remarkable temporal and spectral variability across various timescales  \citep{1985ApJ...299..839C,2017ApJ...844...16H}. In Table \ref{4u_parameters} a brief summary of the system's main characteristics is provided.

4U 0114$+$65 is a non eclipsing system \citep{2015MNRAS.454.4467P} and hosts an X-ray pulsar with a $\sim 2.6$ h spin period, which makes it one of the slowest X-ray pulsars ever known. In order to explain such a slow spin period \cite{1999ApJ...513L..45L} and \cite{2017A&A...606A.145S} suggested that this object could have been born as a magnetar. Long periods of X-ray pulsars might point to high magnetic fields of accreting NS \citep{2013ARep...57..287I}. Other slowly rotating NSs were hypothesized to be magnetars, similar to the case of the slowly rotating NS found in the supernova remnant RCW103 \citep{2016arXiv160704107R,2016arXiv160704264D}. 

Magnetars are NSs that possess very strong magnetic fields, $B\simeq 10^{14}-10^{15}$ G \citep{1992ApJ...392L...9D} and they can be present in HMXRBs \citep{2008ApJ...683.1031B}. Alternatively, \cite{2006A&A...458..513K} proposed that the observed pulse in this system could be explained by the accretion of a structured wind produced by tidally driven oscillations in the B-supergiant star photosphere induced by the closely orbiting NS. 

The pulse period of 4U 0114$+$65 has been changing fast over the years. It was initially estimated to be around 2.78 hours \citep{1992A&A...262L..25F}. Subsequent observations reported shorter periods: in 2000, \cite{2000ApJ...536..450H} found a period of approximately 2.73 hours, followed by \citep{2005A&A...436L..31B} who measured it to be about 2.67 hours. In 2006, \cite{2006MNRAS.367.1457F} obtained a period of roughly 2.65 hours. \cite{2011MNRAS.413.1083W} noted a further decrease in the spin period, from approximately 2.67 hours to 2.63 hours between 2003 and 2008, with a corresponding increase in the NS spin rate of approximately $1.09 \times 10^{-6}\, \text{s}^{-1}$, in \cite{2017A&A...606A.145S} a period of 2.6 h was reported. The pulse period is so long that allows the detailed study of intrapulse morphology. 

Furthermore, the system's long-term behavior exhibits super-orbital modulations and torque reversals, suggesting complex interactions between the NS and the accreting material \citep{2006MNRAS.367.1457F,2006AdSpR..38.2779S}. \citet{2017ApJ...844...16H} proposed that the changing behavior of the NS spin period could be caused by the effect of a transient disk.

In \citet{2017A&A...606A.145S} the spin behavior (duration and fast evolution) of 4U 0114$+$65 was explained within the framework of the quasi-spherical settling accretion theory proposed by \cite{2012MNRAS.420..216S}. In this mode of wind accretion, typically present in sources exhibiting moderate X-ray luminosity (below $\simeq 4\times 10^{36}$ergs$^{-1}$) a convective quasi-spherical shell forms above the NS magnetosphere. The inflow rate of plasma from this shell through the magnetosphere is governed by the cooling mechanisms acting on the hot plasma, primarily driven by Compton and radiative energy losses.

Relative recent observations with \textit{NuSTAR} and \textit{XMM-Newton} have shed new light on the system's activity, revealing features such as Corotating Interaction Region (CIR), whose presence is commonly invoked to explain the periodic variability observed in the UV spectral lines formed in stellar winds of B-type supergiants \citep{2008ApJ...678..408L} as well as the modulation of X-ray emission from O stars \citep{2001A&A...378L..21O, 2014MNRAS.441.2173M}, off-states and potential signatures of Cyclotron Resonant Scattering Features (CRSF) \citep{2023MNRAS.522.3271A} which could be at odds with the magnetar nature of the system.

In this paper, we present a study of 4U 0114$+$65 based on a 200 ks \textit{Chandra} gratings observation allowing us to infer characteristics of the accretion flow along several phases of the spin and orbital period.

\begin{table}
\centering
\caption {Main astrophysical parameters of 4U0114$+$65.}
\begin{adjustbox}{max width=\columnwidth}
\begin{tabular}{rrr}
\hline
Companion & & \\
\hline
Spectral Type & B1Ia & \citet{Reig} \\
T$_{\rm eff}$ (K) & $24000\pm3000$ & \citet{Reig} \\
Radius ($R_{\odot}$) & $37\pm15$ & \citet{Reig} \\
Mass ($M_{\odot}$) & $16\pm5$ & \citet{Reig} \\
& & \\
\hline
System & & \\
\hline
$d$ (kpc) & $4.5 ^{+0.3}_{-0.2}$ & \citet{Bailer-Jones_2021} \\
$M_{V}$ & $-7\pm1$ & \citet{Reig} \\

$E(B-V)$ & $1.24\pm0.02$ & \citet{Reig} \\
$BC$ & $-2.3\pm0.3$ & \citet{Reig} \\
$M_{\rm bol}$ & $-9.3\pm 1.0$ & \citet{Reig} \\
$L_{X} (\times 10^{36} \rm{erg} \ s^{-1}) $ & 0.1 - 1.3 & This work \\
&&\\
$P_{\rm orb}$ (days) & $11.6\pm0.1$ & \cite{grun} \\
$P_{\rm superorb}$ (days) & $30.7\pm0.1$ & \citet{2006MNRAS.367.1457F} \\
$P_{\rm pulse}$ (s) & $9050\pm100$ & This work \\
&&\\
Inclination & $\sim$ 45 $^{\circ}$ & \cite{grun} \& This work\\
Eccentricity& 0.18 $\pm$0.05 &\cite{grun,1985ApJ...299..839C} \\
Angle to the periapsis & 11 $\pm$ 11 $^{\circ}$ & \cite{grun} \\

$v \sin i$ & $96\pm20$ & \citet{Reig} \\
$v_{\infty}$ (km/s) & 1200 & \citet{Reig} \\
&&\\
R$_{\rm BONDI}$(cm) & $\approx$ 10$^{11}$&\\
R$_{\rm COROTATION}$(cm) & $\approx$ 7 $\times$ 10$^{10}$&\\
R$_{\rm ALFVEN}$(cm) & $\approx$ 1.7 $\times$ 10$^{10}$&\\

\hline
\\
\end{tabular}%
\end{adjustbox}
\tablefoot{The velocity of the stellar wind was calculated with a $\beta$-law equation: $v_{wind}=v_\infty(1-\frac{R_*}{a})^\beta$, with the value $\beta=0.8$ (\citet{1999isw..book.....L}), at the site of the NS.$R_{*}$ refers to the companion star radius. Where the R$_{\rm BONDI}$ is \( R_B \approx \frac{2GM}{v_w^2 + v_{\text{orb}}^2} \), R$_{\rm COROTATION}$ \( R_{\text{cor}} \approx \left(\frac{GMP_*}{4\pi^2}\right)^{1/3} \), and R$_{\rm ALFVEN}$ corresponds to the magnetospheric boundary.}

\label{4u_parameters}%
\end{table}%

\begin{figure}
\centering
\subfigure{\includegraphics[trim={0cm 0cm 0cm 10cm},width=1.0\columnwidth]{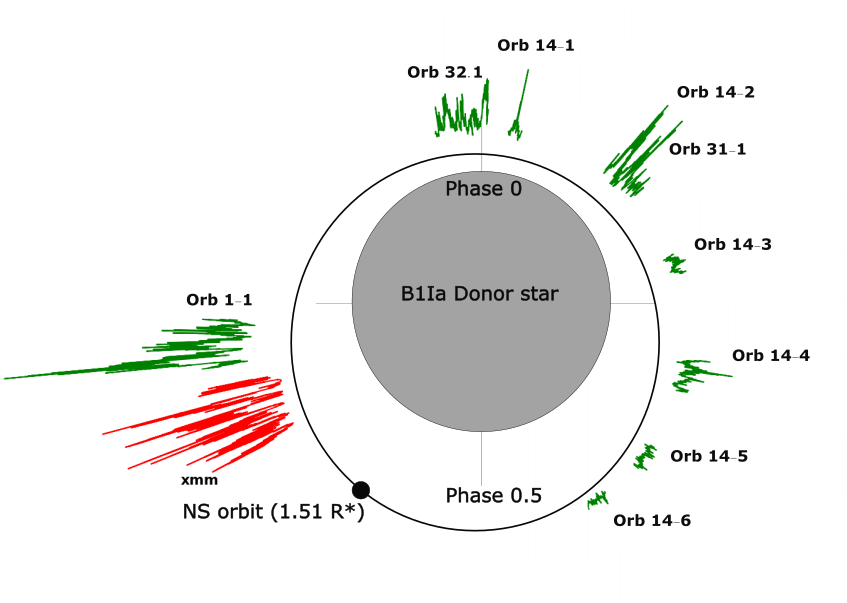}}
\caption{Pole-on sketch of the system showing the relative sizes of the companion star and the NS orbit (NS size not to scale, the observer would be situated in a perpendicular plane with respect to the orbit). The light curves are displayed parallel to the orbit. Green represents the 9 uninterrupted \textit{Chandra} light curves from the dataset analyzed in this paper scaled to the count rate. Red represents a former \textit{XMM-Newton} light curve taken in 2015 scaled accordingly to fit the figure. }
\label{4u_orbita}%
\end{figure}

\section{Observations and analysis}
\label{obs_an}

The data set analyzed consists of nine different \textit{Chandra} observations spread along the NS orbit, taken at different epochs. The \textit{Chandra} telescope, launched by NASA in 1999, employs a system of four nested grazing-incidence mirrors coated with iridium. These mirrors focus X-rays onto detectors with an angular resolution of 0.5 arcseconds. \textit{Chandra} operates over an energy range of $0.1$–$10$ keV, using two primary instruments: the High Resolution Camera (HRC) for imaging and the Advanced CCD Imaging Spectrometer (ACIS) for detailed spectroscopy. Positioned in a highly elliptical orbit, it minimizes Earth interference and enables long-duration observations \citep{Weisskopf_2000}. For this analysis, the ACIS data were utilized.

The log of the observations, the corresponding orbit and order within each orbit, are collected in Table \ref{log}.

The spectra and response (arf and rmf) files were generated using standard procedures with \textsc{ciao} software v4.15, CalDB 4.15, \citep{2006SPIE.6270E..1VF}. The first dispersion orders ($m = \pm 1$) of \textsc{heg} and \textsc{meg} were extracted and combined as described in Section \ref{sec:spectra}. Useful data was considered in the $0.2$-$10$ keV range.

The spectra were analyzed and modeled with the \textsc{isis}\footnote{https://space.mit.edu/cxc/isis/} package. The emission lines were identified using the \textsc{atomdb}\footnote{http://www.atomdb.org/} data base. The orbital modulations present within this data set were analyzed using the \textsc{xraybinaryorbit} python package \citep{Sanjurjo-Ferrin2024} \footnote{https://xragua.github.io/xraybinaryorbit/}.

\begin{table}
\centering
\caption{Observation log.}
\begin{adjustbox}{max width=\columnwidth}
\begin{tabular}{cccccccc}

Obs I.D. & Instrument & Grating & Exposure (ks) & Start date & Orbit & Number \\
\hline
24482 & ACIS-S & HETG & 30.98 & 02/06/21 08:43 & 1 & 1 \\
24481 & ACIS-S & HETG & 9.93 & 03/11/21 14:19 & 14 & 1 \\
26180 & ACIS-S & HETG & 9.83 & 04/11/21 13:42 & 14 & 2 \\
23432 & ACIS-S & HETG & 15.75 & 05/11/21 10:392 & 14 & 3 \\
26177 & ACIS-S & HETG & 24.98 & 06/11/21 09:18 & 14 & 4 \\
26178 & ACIS-S & HETG & 19.81 & 07/11/21 05:12 & 14 & 5 \\
24480 & ACIS-S & HETG & 17.78 & 07/11/21 19:11 & 14 & 6 \\
24479 & ACIS-S & HETG & 28.84 & 20/05/22 15:55 & 31 & 1 \\
24483 & ACIS-S & HETG & 38.51 & 30/05/22 17:20 & 32 & 1 \\
\hline
\end{tabular}
\end{adjustbox}
\label{log}
\end{table}

\section{Timing analysis}
\label{sec:timming}

Within this dataset there are nine different uninterrupted light curves spread across four different orbits. Each light curve will be referred to as the orbit in which it occurred followed by a number in chronological ascending order. As an example, the first light curve will be Orb 1-1 (see Table \ref{log} and \href {https://zenodo.org/records/14645709}{Fig A.1. in the APPENDIX}). The orbital phases were calculated following the ephemeris provided by \cite{2017ApJ...844...16H} (see Eq.2 in their paper). The light curves were extracted in 10\,s bins and then further re-binned to reach signal-to-noise 7 in each bin. Thus, the final light curves have different time resolution ranging between 30 to 200\,s. The pulse period of the NS can be easily discerned by the eye. To characterize the spin period is not straightforward thought (see below).

\begin{figure}
\centering
\subfigure{\includegraphics[trim={0cm 0cm 0cm 0cm},width=1\columnwidth]{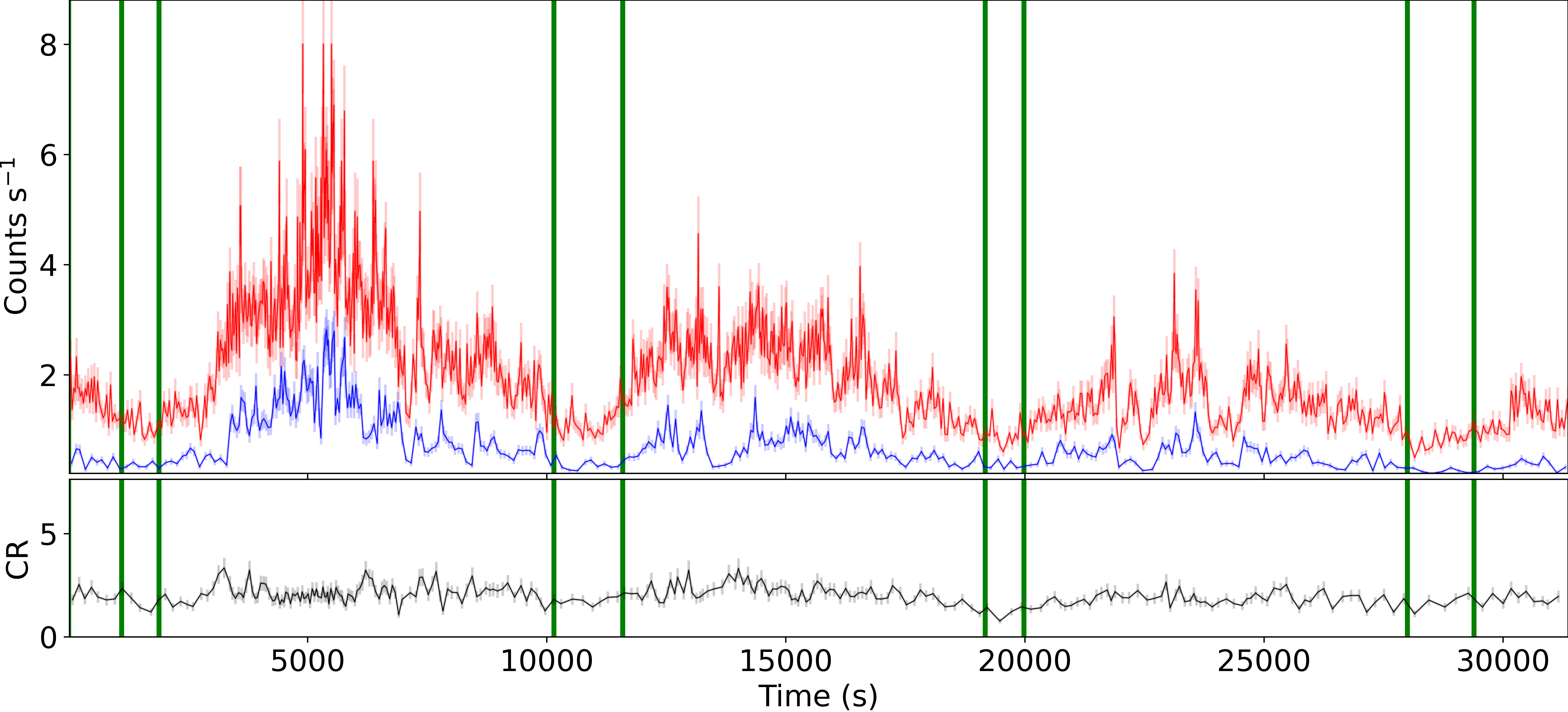}}

\caption{Example of peaks and valleys division on Orb 1-1. Upper panel: High-energy (3–10 keV, red) and low-energy (0.2–3 keV, blue) light curves. The vertical lines indicate the division between peaks and valleys. Lower panel: Color ratio (CR), defined as the high-energy light curve divided by the low-energy light curve.}

\label{fig:obs11lc}%
\end{figure}

To facilitate the analysis of the light curve and perform spectral analysis, we segmented each observation into peaks and valleys. This segmentation was achieved by re-binning the light curve using a weighted mean approach. A valley was identified at each minimum, while a transition to a peak was considered when the count recovery exceeded three times the median difference in counts between bins for each light curve (see Fig.\ref{fig:obs11lc} for an illustrative example).

In Table \ref{tab:lc_stats} we provide comprehensive data on the light curves, including their duration, phase, NS period if detected, pulse fraction, duration of valleys and peaks and weighted mean color ratio (CR).

\begin{table*}
\centering
\caption{Relevant statistical parameters of each observation within this dataset.}
\begin{adjustbox}{max width=\textwidth}
\begin{tabular}{cccccccc}
\hline
Light curves & Duration (s) & Phases & Period (s) & Pulse fraction & Valley Duration (s) & Peak Duration (s) & CR \\
\hline
Orb 1-1 & 31390 & 0.73-0.76 &     9200 $\pm$ 400      & 0.6 $\pm$ 0.6 & 1100 $\pm$ 400 & 8000 $\pm$ 300 & 1.8 $\pm$ 0.8 \\
Orb 14-1 & 10050 & 0.03	0.04 & - & 0.53 $\pm$ 0.05 & 2420 $\pm$ 100 & 4520 & 1.2 $\pm$ 0.9 \\
Orb 14-2 & 9940 & 0.11-0.12 & - & - & 2860 $\pm$ 100 & - & 1.4 $\pm$ 0.9 \\
Orb 14-3 & 15950 & 0.19-0.20 & - & - & 2300 $\pm$ 1000 & 7060 & 1.0 $\pm$ 0.8 \\
Orb 14-4 & 25300 & 0.27-0.29 & 8700 $\pm$ 500     & 0.32 $\pm$ 0.06 & 2040 $\pm$ 120 & 7200 $\pm$ 400 & 1.2 $\pm$ 0.8 \\
Orb 14-5 & 20050 & 0.34-0.36 & - & 0.21 $\pm$ 0.07 & 1600 $\pm$ 300 & 7000 $\pm$ 1900 & 1.1 $\pm$ 0.8 \\
Orb 14-6 & 18010 & 0.40-0.41 & - & 0.21 $\pm$ 0.07   & 1400 $\pm$ 1000 & 6530 & 1.0 $\pm$ 0.7 \\
Orb 31-1 & 29230 & 0.11-0.13 & 9200 $\pm$ 800& 0.47 $\pm$ 0.05 & 1500 $\pm$ 300 & 7300 $\pm$ 600 & 1.5 $\pm$ 0.9 \\
Orb 32-1 & 39020 & 0.99-0.01 & - & 0.40 $\pm$ 0.06 & 1400 $\pm$ 300 & 9000 $\pm$ 2400 & 1.5 $\pm$ 1.2 \\
\hline
\end{tabular}
\end{adjustbox}
\label{tab:lc_stats}
\end{table*}

\subsection{Spin period}

The slow pulsation of the X-ray source can be appreciated at first sight. However the pulse is not consistently detected in every observation, as evidenced by the Lomb-Scargle periodogram \citep{1982ApJ...263..835S}. Specifically, during Orbit 14, the pulse is absent in the first two observations, although this might be caused by their duration's being close to the NS spin period. In observations 5 and 6, the pulse is weakly detected. Furthermore, in Obs 32-1, the pulse appears irregular and interrupted, leading to its non detection in the periodogram.

The pulse fraction is a measure of the amplitude of variability in a periodic or pulsating signal. It can be calculated as:
\[
\text{PF} = \frac{I_{\text{max}} - I_{\text{min}}}{I_{\text{max}} + I_{\text{min}}}
\]
where $I_{\rm max}$ and $I_{\rm min}$ are the maximum and minimum intensity observed during the pulse, respectively. 

Within this source, the pulse fraction varies considerably across observations (Table \ref{tab:lc_stats}), ranging from 0.6 to 0.2. There is a strong positive correlation ($r=0.9$) between the evolution of the pulse fraction through the orbit and a sinusoidal orbital evolution. 


The source exhibits secular spin up with periods of reversal, as documented in \citet{Hu_2017} and \citet{2023MNRAS.522.3271A}. To accurately calculate the spin period, we utilized a Lomb-Scargle periodogram on the observations that exhibit a pulse fraction higher than 0.3, where the pulse could be detected: Orbit 1-1, Orbit 14-4, and Orbit 31-1. The weighted mean period, along with its error, was determined to be 9050 $\pm$ 100 seconds. The source is thus continuing the spin-up trend previously reported.

\subsection{NS pulse shape}
\label{nspulses}

A remarkable characteristic of this source is that each pulse is different to each other. Moreover, Orb 1-1, Orb 14-3 and Orb 32-1 show a decreasing evolution of the peak count rate, both in the high and low energy light curves.

To further investigate this issue we selected all the peaks within this dataset lasting more than 4000 s and calculated the weighted mean of the count rates for the high energy light curve (whc, 3-10 keV), the low energy light curve (wlc, 0.3-3 keV) and the hardness ratio (HR) defined as $\frac{whc-wlc}{whc+wlc}$ and used a \texttt{k-means} algorithm to classify them. \texttt{k-means} \citep{MacQueen1967SomeMF} is a popular clustering algorithm used in machine learning to partition a dataset into a predetermined number of clusters. It works by iteratively assigning data points to the nearest cluster centroid and then updating the centroids based on the mean of the points assigned to each cluster. 

The best result obtained was for 4 groups with a \textit{silhouette score} of 0.52. The silhouette score \citep{ROUSSEEUW198753} measures how similar an object is to its own cluster compared to other clusters, with values ranging from -1 to 1. A score of -1 indicates no separation between clusters, while a score of 1 suggests a perfect separation. A score of 0.52 indicates a reasonably good clustering result, suggesting that the data points are appropriately grouped into distinct clusters. Based on their characteristics, we labeled each of the groups as Bright, Intermediate, Faint, and Absorbed (Fig. \href {https://zenodo.org/records/14645709}{Fig A.1. in the APPENDIX} and Table \ref{tab:clusters}).

\begin{table}
\centering
\caption{Cluster mean and standard deviation values for the \texttt{k-means} classified peaks.}
\label{tab:clusters}
\begin{tabular}{lccc}
\hline
\textbf{Spectra} & \textbf{High lc} & \textbf{Low lc} & \textbf{HR} \\
\hline
Bright       & $2.0 \pm 0.2$ & $0.7 \pm 0.1$ & $0.50 \pm 0.01$ \\
Intermediate & $1.2 \pm 0.1$ & $0.4 \pm 0.1$ & $0.50 \pm 0.04$ \\
Faint        & $0.6 \pm 0.1$ & $0.30 \pm 0.04$ & $0.40 \pm 0.03$ \\
Absorbed     & $1.5 \pm 1.2$ & $0.20 \pm 0.02$ & $0.60 \pm 0.05$ \\
\hline
\end{tabular}
\end{table}

\begin{figure}
\centering
\subfigure{\includegraphics[trim={0cm 0cm 0cm 0cm},width=1\columnwidth]{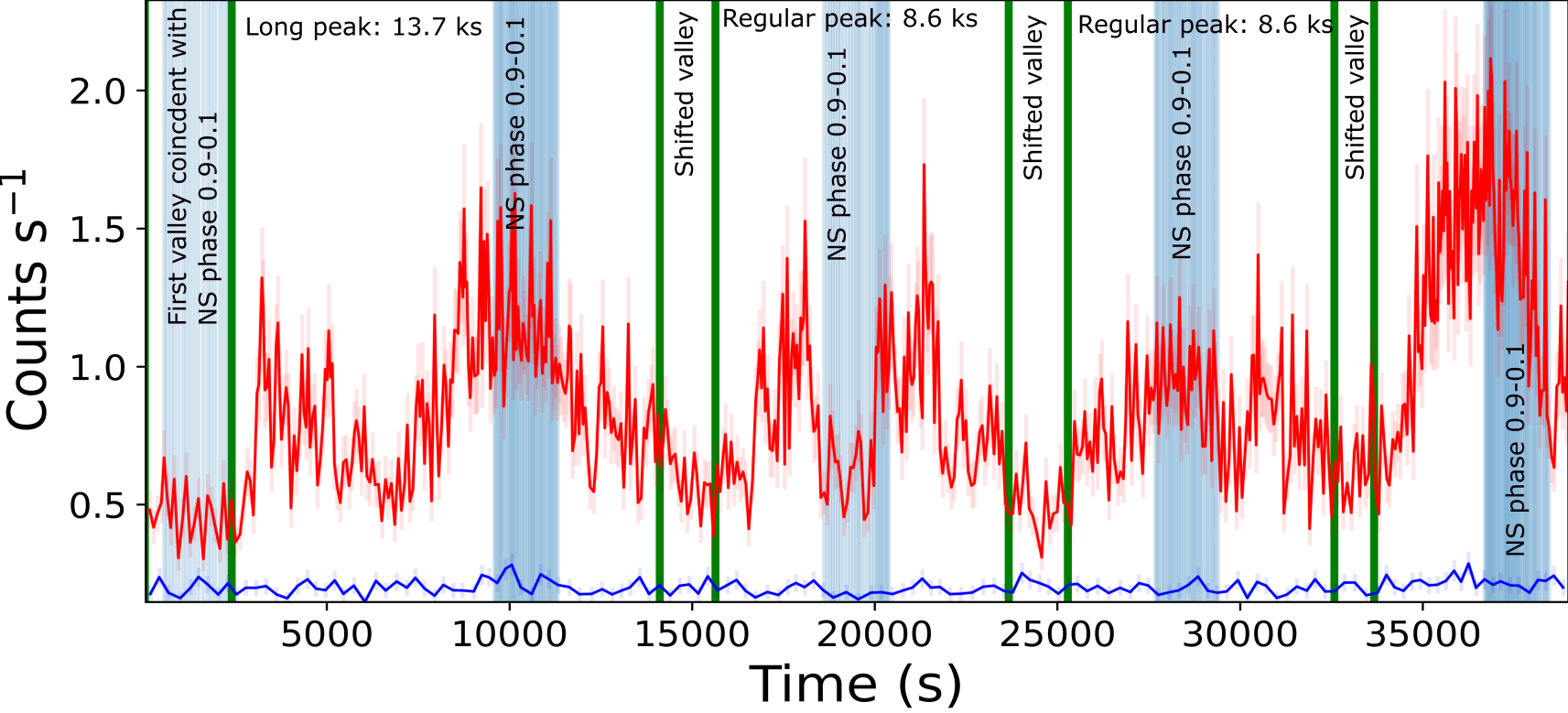}}
\caption{Orb 32-1 light curve. Green vertical lines divide peaks from valleys. NS phase $0.9-0.1$ is shadowed in blue. }
\label{lc_orb_32}%
\end{figure}

In the Orb 32-1 light curve (Fig. \ref{lc_orb_32}), 
the first peak lasts 13.7 ks, approximately $\times 1.5 $ the duration of the rest of the other peaks ($\sim 9.0\pm 0.7$ ks on average). The subsequent peaks recover the $\sim$ 9ks duration. However a dramatic change occurs: the first valley is placed in the NS corresponding valley phase ($0.9-0.1$). But, after the first 13 ks peak, the NS phase is shifted by 0.5 so that now, peaks are where valleys should be.

In \citet{2017A&A...606A.145S} a low luminosity episode, making one of the NS spin valleys longer than expected was interpreted as a substantial departure from the spherical accretion symmetry. The Bondi radius is an idealization, which could change by a factor 2 or 3 in different directions, as the stellar wind velocity changes on the characteristic scale of $10^{11}$ cm, which is comparable to $R_{\rm B}$ (Table \ref{4u_parameters}). The stellar wind could then be momentarily captured in a non-spherical way. Since the magnetospheric radius and the Bondi radius differ only by a factor of a few, the captured matter can reach the magnetosphere non spherically, thereby enabling a switching between the magnetic poles of the NS. Our hypothesis is that a pole switching could cause a longer than normal peak and an immediate recover to the usual values, maintaining, from now on, a 0.5 NS phase shift.

\begin{figure}[h!]
\centering
\subfigure{\includegraphics[trim={0cm 0cm 0cm 0cm},width=1\columnwidth]{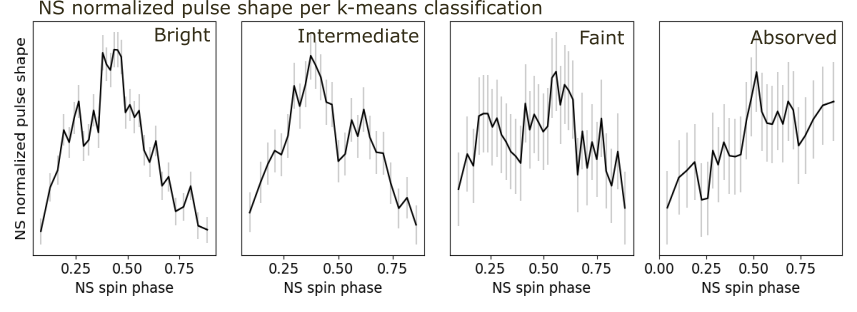}}
\caption{Normalized NS spin pulse shapes. The averaged pulses of the four distinct orbits included in this dataset are shown in the top panels, while the averaged pulses for the k-means classifications are displayed in the bottom panels. To compute the average pulse shape, all pulses were normalized beforehand to ensure that each contributed equally to the final result.}
\label{fig:pulse_shape}%
\end{figure}

To analyze the pulse profile, we folded the peaks identified within the k-means classification for each of the four clusters (see Fig. \ref{fig:pulse_shape}). Prior to folding, the peaks were normalized to ensure that each peak contributed equally to the average pulse profile. The folded pulse for the Bright group displays a broad maximum but is skewed to the left, with higher intensity observed at lower NS spin phases. The folded pulse for the Intermediate group indicates a subtle double-peaked structure, with the left peak appearing more pronounced than the right. In both cases, the maximum is situated around NS phase 0.4. In contrast, the Faint and Absorbed groups exhibit irregular profiles with large error bars, precluding detailed analysis. \citet{2003A&A.404.1023B} noted the scarcity of double-peaked accretion-powered pulsars with respect to what is expected from theoretical models, attributing this scarcity to the likely alignment of the magnetic and rotational axes, which would increase the probability of single peak pulse profiles. When compared to the pulse profiles calculated using \textit{XMM-Newton} data by \citet{2017A&A...606A.145S} (see their Fig. 4), the pulses in this dataset share characteristics with those observed in the first part of their observation (T1).

\subsection{Spikes}

The very slow cadence of the NS pulsations allows us to appreciate a lot of interesting details within each peak, revealing the complexity of the accretion processes. In this Section we want to focus on the flickering variability exhibited by this source, specially within the peaks, were some spikes can be observed. 

There are two main challenges in the spike detection. On one hand we have to distinguish them from random noise. On the other hand, we need to retain a sufficient time resolution to obtain reliable results. Unfortunately, within this dataset the low energy light curve is very suppressed, so we will focus solely on the spikes detected in the high energy light curve. 

To distinguish real spikes from the noise, we conducted a comparative analysis using a semi-supervised dip-detection algorithm based on anomaly detection with autoencoders. Autoencoders are known for their effectiveness in anomaly detection tasks, as they excel at learning and reconstructing normal data patterns \citep{2020arXiv200305991B}. By leveraging the reconstruction error, they offer a robust mechanism to identify data points that deviate from the norm. 

In our study, we generated synthetic light curves by de-trending the original data and creating a random time series with the same statistical properties as the original light curve. The spikes detected in these synthetic light curves were then used as a training set for our autoencoder model. Spikes observed in the real light curve that deviated significantly from those detected in the synthetic data were identified and preserved as genuine spikes likely caused by astrophysical phenomena. Additionally, we calculated the signal-to-noise ratio (SNR) based on the mean uncertainty of the signal around each spike, and spikes with an SNR greater than five were also accepted.

We will define the prominence of each spike as their increase in count rate with respect to their baseline divided by the baseline. Thus, the prominence represents the increase in percentage of the count rate.

\begin{figure}[h]
\centering
\subfigure{\includegraphics[trim={0cm 0cm 0cm 0cm},width=1\columnwidth]{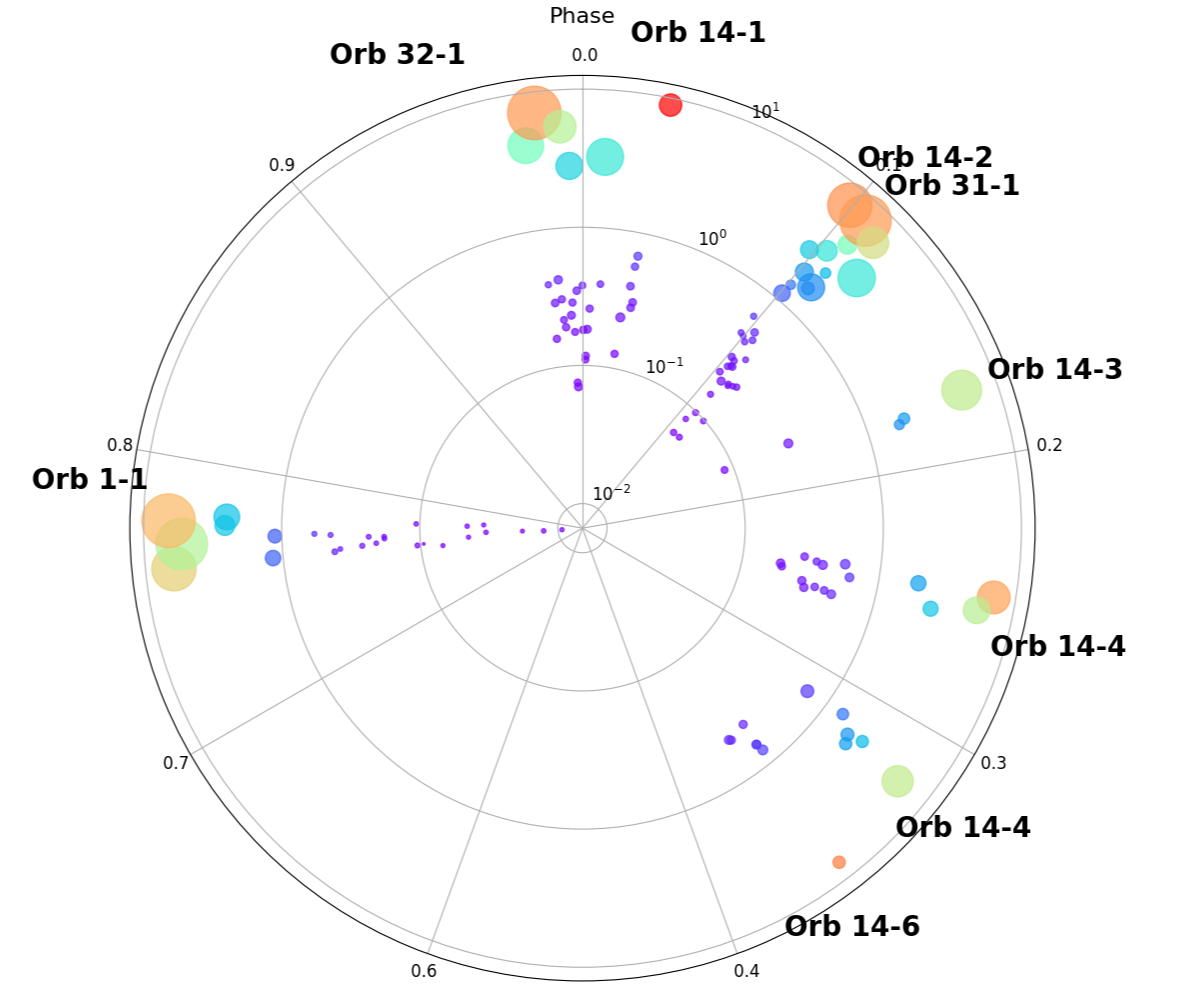}}
\caption{Detected spikes at their respective orbital phases. The radial axis represents the prominence. The size of the bubbles is proportional to their duration. The color gradient is used to support visually the prominence gradient. }
\label{fig:dips_orbit}%
\end{figure}

\begin{figure*}[h]
\centering
\subfigure{\includegraphics[trim={0cm 0cm 0cm 0cm},width=1\textwidth]{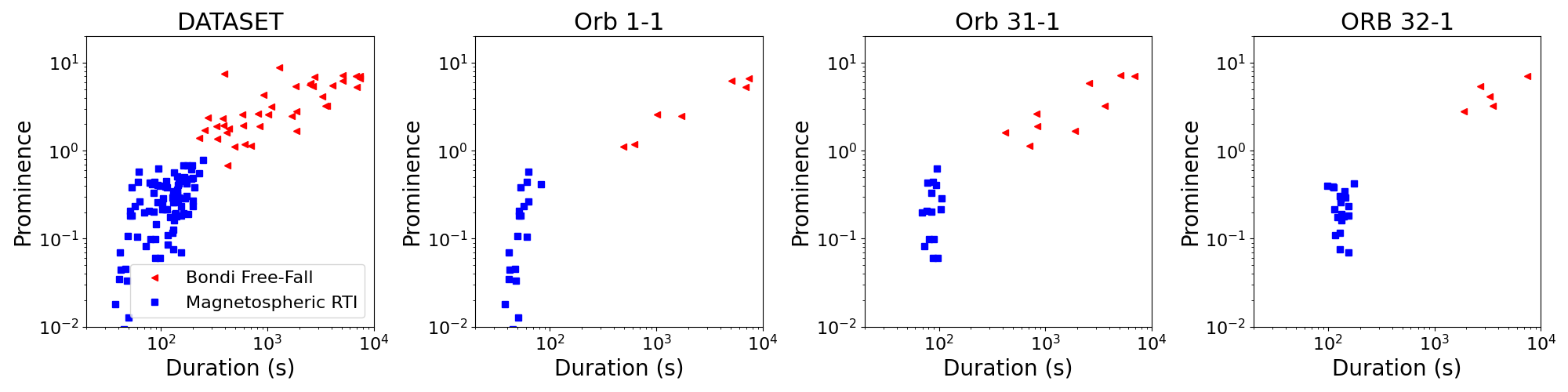}}
\caption{Prominence vs duration of spikes detected in this dataset. Blue squares and red triangles represent the two separated clusters obtained by a \texttt{gaussian mixture} algorithm. The leftmost panel provides the results obtained for all the observations combined, while the three panels on the right display the results for the three individual observations that produced the better results in number of spikes detected and thus, in clustering.}
\label{fig:spikes_orbit_gm}%
\end{figure*}

After obtaining the real spikes we applied a \texttt{gaussian mixture} clustering algorithm. Unlike the k-means algorithm, which assumes clusters of spherical shape, \texttt{gaussian mixture} provides a more flexible clustering approach by modeling the data as a mixture of multiple Gaussian distributions, allowing for elliptical cluster shapes and different cluster sizes \citep{Reynolds2009}. This approach led to better results in this particular case, achieving a silhouette score greater than 0.6, which indicates the presence of two well-separated clusters. A higher silhouette score (between 0.7-0.8) was achieved by clustering each observation's spikes individually, pointing to a difference in the spikes duration and prominence through the different observations (see Fig. \ref{fig:spikes_orbit_gm}).

\begin{figure}
\centering
\subfigure{\includegraphics[trim={0cm 1cm 0cm 1cm},width=0.8\columnwidth]{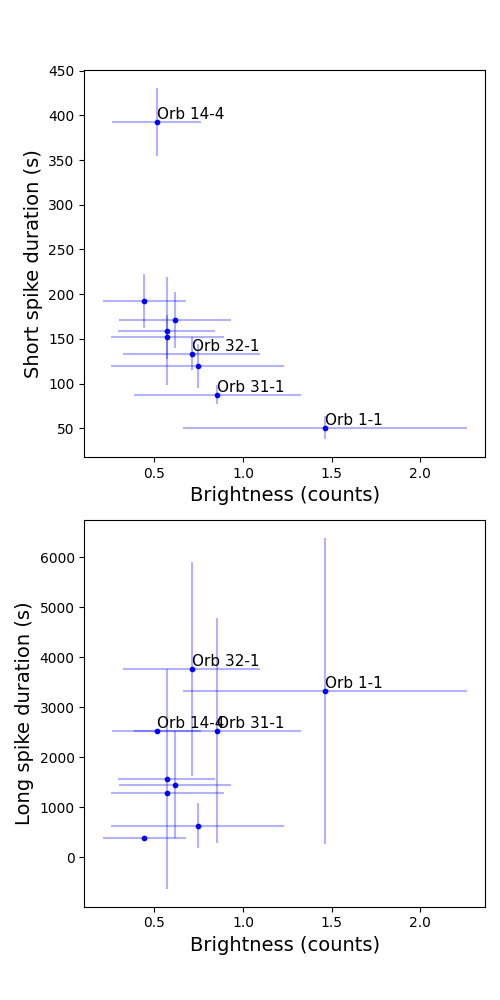}}
\caption{The plots show the average spike duration for each of the two clusters (with the upper panel displaying the short spikes and the lower panel displaying the long spikes) as predicted by the \texttt{Gaussian Mixture} algorithm for each observation. The x axis represents the brightness of each observation, measured as the error-weighted count average (whc).
}
\label{Fig:duration_brightness}%
\end{figure}

Two distinct groups of spikes are observed within this dataset, clearly distinguished by their duration and prominence. We refer to them as short and long spikes based on their duration. The short spikes, within each individual observation, exhibit a narrow range of durations (see the three right panels in Fig. \ref{fig:spikes_orbit_gm}). Notably, the duration of these short spikes is inversely proportional to the brightness of the source for each observation, as measured by the weighted mean number of counts. In contrast, no specific trend is observed for the long-duration spikes (see Fig. \ref{Fig:duration_brightness}). Similar flaring activity has been attributed to the development of magnetospheric instability in NS in other HMXRB, such as Supergiant Fast X-ray Transients \citep{2019MNRAS.487..420S}.

\section{Spectral analysis}
\label{sec:spectra}

\begin{figure*}[ht]
\centering
\subfigure{\includegraphics[trim={0cm 0cm 0cm 0cm},width=0.9\textwidth]{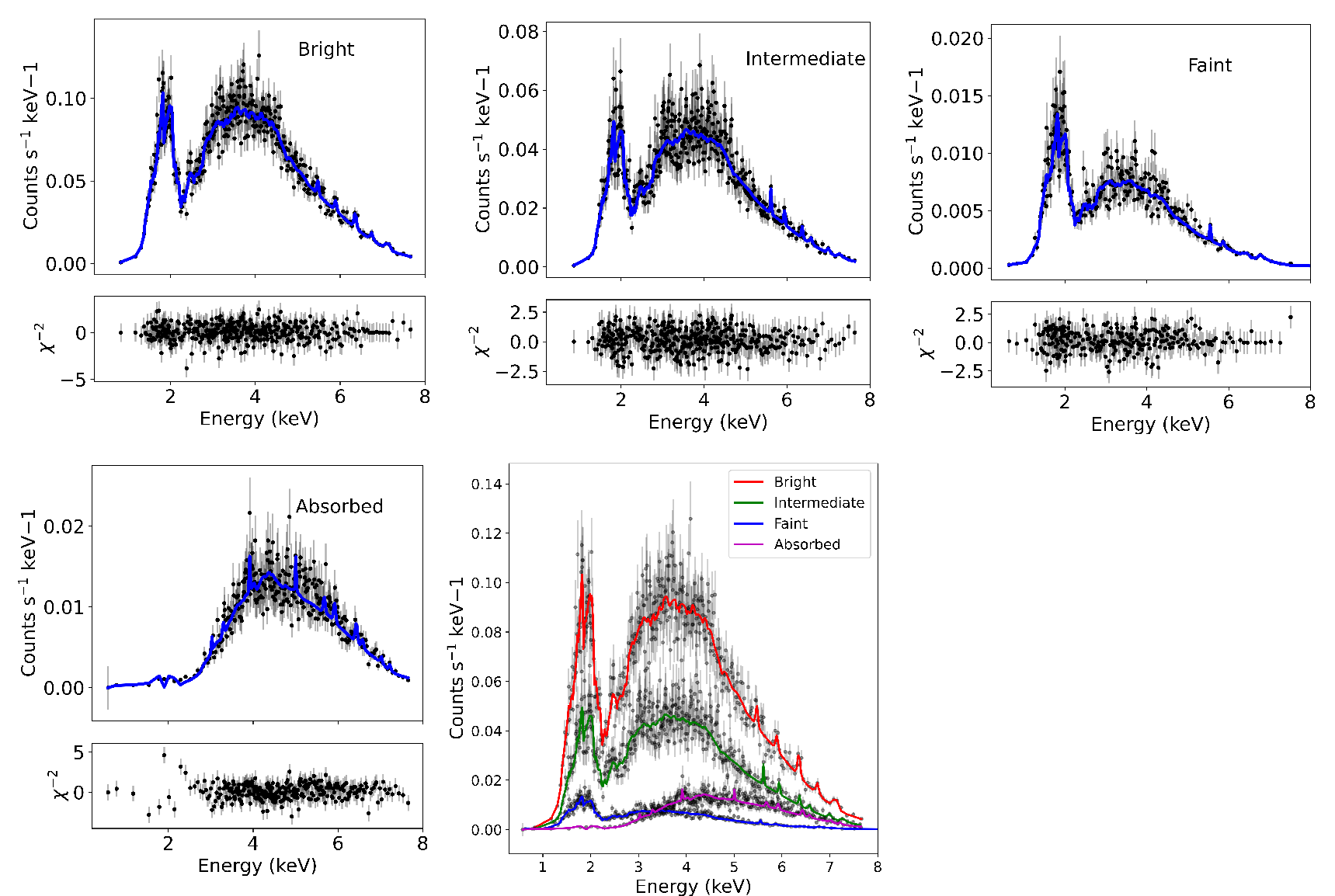}}
\caption{Upper row (left to right): Spectra, model, and $\chi^2$ values for Bright, Intermediate and Faint. Lower row (left to right): Spectra, model, and $\chi^2$ values for Absorbed and all spectra displayed for comparison.}
\label{fig:kmeans_spectra}%
\end{figure*}

We used two different criteria to perform the spectral analysis. On one hand, we analyze the combined spectra for each one of the four orbits in this dataset. To achieve this we isolated and combined the peaks within each observation.

These spectra are Orb 1, Orb 14, Orb 31 and Orb 32. However, the limited count rate prevents us from conducting the same exercise with each orbit for the valleys. Thus, all valleys were combined in one single spectrum which will be referred as valley combined. This classification will be referred to as orbital spectra. The best fit for each spectra is represented in \href {https://zenodo.org/records/14645709}{Fig B.1. in the APPENDIX} and the best fit parameters collected on Table \ref{tab:model_orbital} (see below for the description of the model used).

On the other hand, we combined the spectra based on the peak morphology grouped as suggested by the \texttt{k-means} algorithm (see subsection \ref{nspulses} and Table \ref{tab:clusters}). We will refer to them as Bright, Intermediate, Faint and Absorbed. This classification will be referred to as \texttt{k-means} spectra. The best fit for each spectra is represented in Fig. \ref {fig:kmeans_spectra} and the best fit parameters collected on Table \ref{tab:model_kmeans}.

\begin{table*}
\centering
\caption{Spectral model parameters for the Orbital spectra.}
\label{tab:model_orbital}
\begin{adjustbox}{max width=0.99\textwidth}
\begin{tabular}{ccccccccccc}
\hline

Spectra & $\chi^2$ & ${N_{\rm H}}$ & $K_{\rm bmc}$ & $kT_{\rm bmc}$ & $\alpha$ & $L^{\rm bmc}_{\rm 0.3-10\rm~keV}$ & $R_{\rm bmc}$ & $K_{\rm bb}$ &$L^{\rm bb}_{\rm 0.3-10\rm~keV}$ \\\\

 &  	&	 ($\times10^{22}$ cm$^{-2}$) 	&($\times10^{-3}$) &	(keV)&	 & ($\times$10$^{34}$ erg s$^{-1}$) &  (km) &    & ($\times$10$^{34}$ erg  s$^{-1}$) \\\\

\hline
Orb 1	&	1.0	&	1.9 $^\pm$ 0.1 & 26 $^{+3}_{-4}$ & 1.7 $\pm$ 0.1 & 0.018 $^{+0.004}_{-0.003}$ & 46 $^{+6}_{-8}$ & 1.4 $\pm$ 0.1 & 0.035 $^{+0.014}_{-0.013}$ & 9 $\pm$ 4 \\\\

Orb 14	&	1.3	&	1.3 $\pm$ 0.1 & 0.25 $\pm$ 0.01 & 1.40 $^{+0.03}_{-0.05}$ & 4.0 $^{+0.0}_{-0.8}$ & 4.9 $^{+0.1}_{-0.2}$ & 0.71 $\pm$0.01 & 0.027 $\pm$ 0.006 & 6.8 $\pm$ 1.6 \\\\

Orb 31	&	1.2	&	2.1 $\pm$ 0.2 & 11.0 $^{+0.4}_{-0.5}$ & 1.50 $\pm$ 0.05 & 0.01 $^{+0.01}_{-0.00}$ & 23 $\pm$ 1 & 1.30 $^{+0.02}_{-0.03}$ & 0.029 $\pm$ 0.014 & 7 $\pm$ 4 \\\\

Orb 32 &	1.5	&	23 $^{+3}_{-2}$ & $\sim$ 1.6  & 1.8 $^{+0.2}_{-0.5}$ & 4.0 $^{+0.0}_{-3.3}$ & $\sim$22  & $\sim$0.91  & 0.04 $^{+0.01}_{-0.04}$ & 9 $^{+3}_{-9}$ \\\\

Valley combined	&	1.2	&	1.5 $^{+0.5}_{-0.4}$ & 0.11 $^{+0.02}_{-0.01}$ & 1.5 $^{+0.2}_{-0.3}$ & 4.0 $^{+0.0}_{-3.1}$ & 2.0 $^{+0.4}_{-0.2}$ & 0.37 $^{+0.03}_{-0.02}$ & 0.012 $\pm$ 0.008 & 3.1 $\pm$ 2.0 \\\\

\hline
\end{tabular}
\end{adjustbox}
\end{table*}

\begin{table*}
\centering
\caption{Spectral model parameters for the \texttt{k-means} spectra.}
\label{tab:model_kmeans}
\begin{adjustbox}{max width=0.99\textwidth}
\begin{tabular}{ccccccccccc}
\hline

Spectra & $\chi^2$ & $N_{\rm H}$ & $K_{\rm bmc}$ & $kT_{\rm bmc}$ & $\alpha$ & $L^{\rm bmc}_{\rm 0.3-10\rm~keV}$ & $R_{\rm bmc}$ & $K_{\rm bb}$ & $L^{\rm bb}_{\rm 0.3-10\rm~keV}$ \\\\

 &  	&	 ($\times10^{22}$ cm$^{-2}$) 	&($\times10^{-3}$) &	(keV)&	 & ($\times$10$^{34}$ erg s$^{-1}$) &  (km) &    & ($\times$10$^{34}$ erg s$^{-1}$) \\\\
\hline
Bright	      &	1.0	&	2.1 $\pm$ 0.1 & 30$\pm$ 1 & 1.50 $\pm$ 0.03 & 0.01 $^{+0.01}_{-0.00}$ & 58.0 $^{+1.5}_{-1.4}$ & 2.0 $\pm$ 0.1 &0.056 $\pm$ 0.024 & 14 $\pm$ 6 \\\\

Intermediate &	1.3	&	2.0 $^{+0.2}_{-0.1}$ & 16 $^{+1}_{-6}$ & 1.60 $^{+0.06}_{-0.13}$ & 0.01 $^{+0.04}_{-0.00}$ & 26 $^{+1}_{-10}$ & 1.20 $^{+0.02}_{-0.20}$ & 0.03 $\pm$ 0.01 & 8 $\pm$ 3\\\\

Faint        &	1.2	&	1.1 $\pm$ 0.2 & 0.190 $^{+0.004}_{-0.007}$ & 1.3 $\pm$ 0.1 & 4 $^{+0}_{-2}$ & 3.9 $\pm$ 0.1  & 0.7 $\pm$ 0.1 & 0.022 $\pm$ 0.007 & 5.7 $\pm$ 1.7 \\\\

Absorbed     &	1.5	&	19.0 $^{+1.0}_{-0.4}$ & 1.40 $^{+0.03}_{-0.10}$ & 1.9 $\pm$ 0.2 & 4.0 $^{+0.0}_{-2.5}$ & 17.0 $^{+0.3}_{-1.3}$ & 0.66 $^{+0.01}_{-0.03}$ & 0.02 $^{+0.03}_{-0.02}$ & 5 $^{+7}_{-5}$ \\\\

\hline
\end{tabular}
\end{adjustbox}
\end{table*}

The continuum has been described using the Bulk Motion Comptonization (\texttt{bmc}) model. \texttt{bmc} is an analytical model which describes the Comptonization of soft seed photons by matter undergoing relativistic bulk motion \citep{1997ApJ...487..834T}. The model parameters are the characteristic black body temperature of the soft photon source, a spectral (energy) index ($\alpha$), and an illumination parameter characterizing the fractional illumination of the bulk motion flow by the thermal photon source. 

The absorption is modeled by the Tuebingen-Boulder interstellar medium (ISM) absorption model \texttt{Tbnew}. This model calculates the cross section for X-ray absorption by the ISM as the sum of the cross sections due to the gas-phase ISM, the grain-phase ISM, and the molecules in the ISM \citep{2000ApJ...542..914W}. The detected emission lines were modeled as gaussians. Their parameters are collected in Table \href {https://zenodo.org/records/14645709}{Fig C.1. in the APPENDIX}.

With the aim of maintaining consistency and comparability, in our first attempt we employed the same model described in \citet{2017A&A...606A.145S}. However, for the present dataset, the partial covering was not needed and a soft excess could be observed at low energies. It was successfully fitted with an absorbed \texttt{blackbody}. The temperature and absorption of this soft excess were $0.04-0.07$ keV and a $0.5-0.8 \times10^{22}$ cm$^{-2}$, respectively, in every case. Thus we decided to fix the temperature to the more frequent value (0.06 keV) and fix the absorption to the interstellar value towards the source (0.8 $\times10^{22}$ cm$^{-2}$).

The complete model is described by the following equation:

\begin{equation} \label{eq:model}
\begin{aligned}[b]
F(E) & = \exp(-N_{H}\sigma(E))\times \left(BMC(E)+\sum_{i=1}^{6}(G_{i}) \right )\\
& \quad + \exp(-N_{H}^{\rm ISM}\sigma(E))\times BBody
\end{aligned}
\end{equation}

where $G_{i}$  represent the gaussians describing the emission lines. Both the orbital and k-mean averages were fit with the same model. 

The radius of the source emitting the seed soft photons, which are subsequently comptonised, can be estimated assuming that the source is radiating as a \texttt{blackbody} of area $\pi R_{W}^{2}$ \citep{2004A&A...423..301T}.

\begin{equation}
\label{radius}
R_{W}=0.6\sqrt{L_{34}}(kT)^{-2} [\rm km]
\end{equation}

where $L_{34}$ is the luminosity of the \texttt{bmc} component in units of 10$^{34}$ergs s$^{-1}$ and $kT$ is the temperature of the seed soft photons. In Tables \ref{tab:model_orbital} and \ref{tab:model_kmeans} we show the best fit parameters for each spectra.

The Orb 32 spectra and the Absorbed spectra both exhibit a high absorption column. 
These spectra where taken in orbital phases close to $\phi \sim 0$.  In addition to the six  lines detected in all other spectra, they exhibit numerous emission lines that are not observed elsewhere in this dataset.

\begin{figure*}
\centering
\subfigure{\includegraphics[trim={0cm 2cm 0cm 0cm},width=0.9\textwidth]{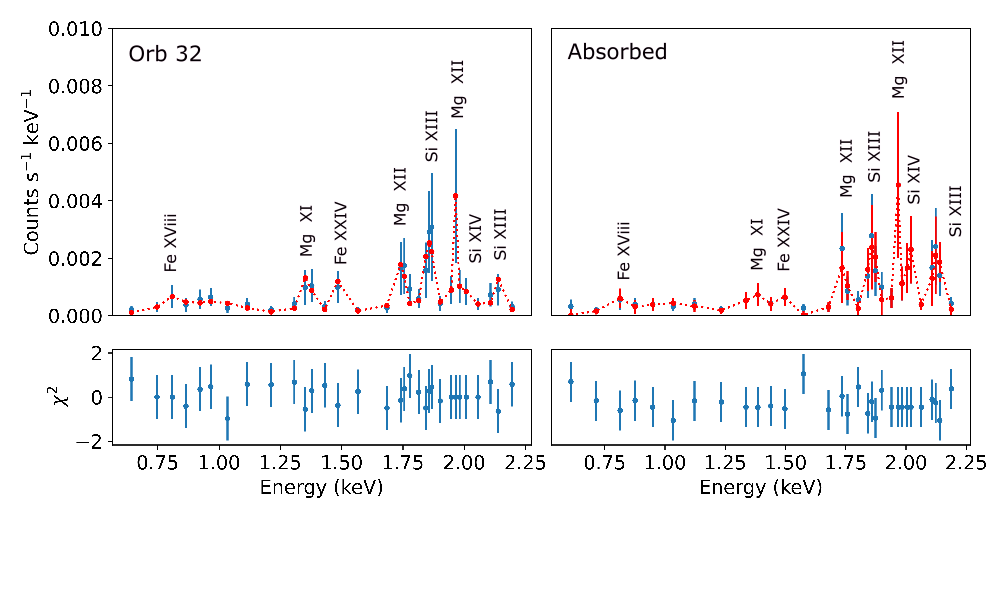}}
\caption{Emission lines observed in the low energy range within the spectra Orb 32 and the absorbed spectra.}
\label{fig:abs_lines}%
\end{figure*}

Particularly intriguing is the presence of a bump near $\sim 2$ keV of these two same spectra. To fit our spectra, depending on the number of counts and the quality of the data, we applied a minimum signal to noise ratio of $5$–$15$ per bin. Upon closer examination, reducing the data binning to a minimum signal to noise ratio of 2.5 per bin revealed that the bump is formed by several emission lines. This change in the signal to noise ratio for the binning in the $0.6$–$2.2$~keV energy range corresponds to a significant increase in resolution, from $R = 4-9$ to $R = 25$–$34$ in terms of energy. This enhancement is well within the \textsc{heg} and \textsc{meg} capabilities of the \textit{Chandra} ACIS instrument (see \cite{Weisskopf_2000} for detailed information). The result is shown in Fig. \ref{fig:abs_lines}. To fit this section of the spectra, we thawed the \texttt{bbody} parameters and add gaussians to fit the detected emission lines. Interestingly, the \texttt{bbody} parameters were compatible with those obtained in the rest of the spectra. Thus, we finally set the temperature and absorption column to 0.06 keV and $0.8\times10^{22}$ cm$^{-2}$ respectively) and let the gaussians to vary freely. These line parameters are collected in Table \href {https://zenodo.org/records/14645709}{Fig D.1. in the APPENDIX}. The emission lines detected on the high energy continuum are reported in Table \href {https://zenodo.org/records/14645709}{Fig E.1. in the APPENDIX}. See eq. \ref{eq:model_absorbed} and eq. \ref{eq:model} for a comparison.

\begin{equation} \label{eq:model_absorbed}
\begin{aligned}[b]
F(E) & = \exp(-N_{H}\sigma(E))\times \left(BMC(E)+\sum_{i=1}^{6+4}(G_{i})\right )\\
& \quad + \exp(-N_{H}^{\rm ISM}\sigma(E))\times \left(BBody\times +\sum_{i=1}^{8}(G_{i}) \right )\\
\end{aligned}
\end{equation}

\section{Discussion}

{\bf\textit{The circumstellar absorption.}} 

The absorption column traversed by the radiation emitted by the compact object on its way to the observer depends on the wind density it trespasses in its path towards the observer. This path depends on the orbital phase, the orbital separation, the eccentricity, inclination and argument of the periapsis. To estimate the density we use the wind structure predicted by the CAK models adopting an acceleration parameter beta $\beta=0.8$ \citep{1975ApJ...195..157C}. 

Since the system has an inclination and eccentricity different from 0, the circumstellar absorption column varies through the orbit, being much higher in phases close to $\phi=0$. This fact explains the high absorption observed in Orb 32 or in the Absorbed spectra, as they were taken during and around $\phi=0$  (Table \ref{tab:model_orbital} and \ref{tab:model_kmeans}).

By comparing the evolution of $N_{\rm H}$ through the orbit with the expected theoretical values we can constrain the inclination and mass-loss rate of the system. In doing so, we will assume that the stellar wind is unionized and spherically distributed. Even when this is an oversimplification of the real situation, this analysis can provide insight into the system configuration.

Utilizing the N$_{H}$ collected in Table \ref{tab:model_orbital} for the Orbital spectra Orb 1, Orb 31 and Orb 32 and the orbital parameters summarized in Table \ref{4u_parameters}, we will obtain an estimation on the mass-loss rate and inclination of the system. Orb 14 comprises several different orbital phases and thus, this data is not suitable for this calculation, and the same problem holds the \texttt{K-means} spectra. 

The complexity of the model prevents us from utilizing a classical least-squares approach; hence, the data was fitted using a Particle Swarm Optimization (PSO) algorithm. In PSO, each particle represents a potential solution to an optimization problem. The particles move through the solution space to find the optimal parameter configuration \citep{10.1162/EVCO_r_00180}.

The best fit ($r=0.9$) was achieved for an inclination of $40\pm3^{\circ}$ and a lower limit for the mass-loss rate of $8.6 \pm 1.7 \times 10^{-7}$ $M_{\sun}$ yr$^{-1}$ (it must be taken into account that the companion's wind could be partly ionized). This result is in agreement with a long term study of the optical spectra \citep{2016A&A...590A.122R} were a strong H$\alpha$ emission was found, showing sometimes a P Cygni profile, therefore, pointing to a substantial wind  density ($\dot M\sim 10^{-6}- 10^{-7} M_{\odot} {\rm yr}^{-1} $).

\begin{figure}
\centering
\subfigure{\includegraphics[trim={0cm 0cm 0cm 0cm},width=0.9\columnwidth]{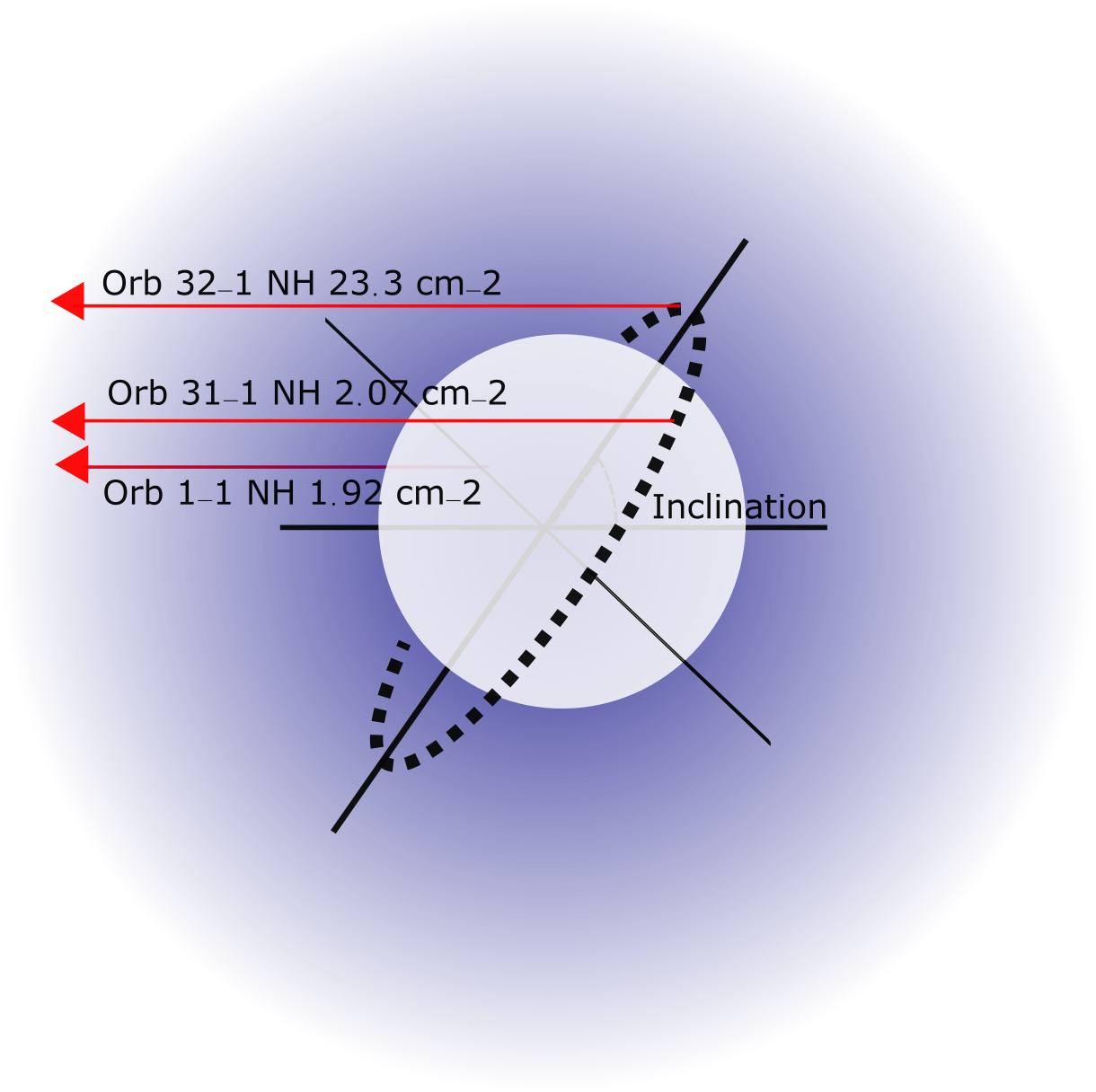}}
\caption{Representation of the system. The dotted circle represents the NS orbit and the white circle the companion star. Red arrows represent the path that the emitting radiation transverses in its path to the observer. This path is different in length and wind density depending on the orbital phase in which the radiation was emitted, provided that the system has an inclination and eccentricity different than zero. The mass-loss rate and inclination could be inferred by the differences found within the $N_{\rm H}$ in three different orbital phases, corresponding to Orb 1-1, Orb 31-1 and Orb 32-1 (see Table \ref{tab:model_orbital} for the $N_{\rm H}$ values and \ref{tab:lc_stats} for the orbital phases).}. 
\label{fig:psonhfit}%
\end{figure}

\begin{figure}
\centering
\subfigure{\includegraphics[trim={0cm 0cm 0cm 0cm},width=0.9\columnwidth]{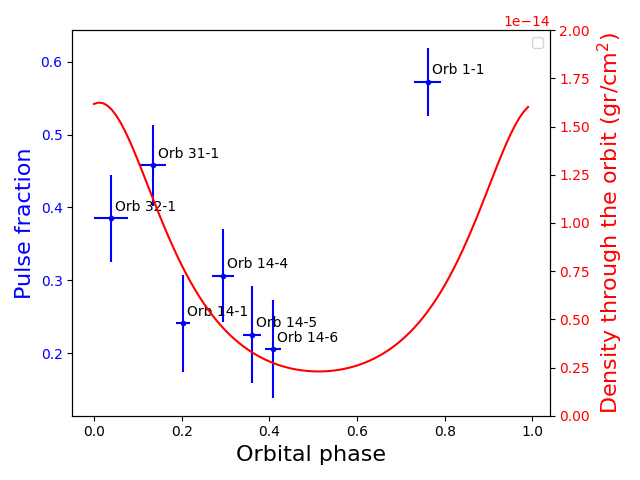}}
\caption{Density variation encountered by the NS as it travels around the eccentric orbit (red continuous line). The measured pulse fraction (Table \ref{tab:lc_stats}) might be correlated with the density of the NS environment. }
\label{fig:psodenfit}%
\end{figure}

A visualization of the system is shown in Fig. \ref{fig:psonhfit}. Due to the system's non-zero eccentricity, the density of the stellar wind encountered by the NS varies through the orbit. This variation arises because the distance between the compact object and the companion star changes as the object moves along its elliptical trajectory. Interestingly the observed pulse fraction varies in a very similar manner as the density trough the orbit expected for the best fit parameters (see Fig. \ref{fig:psodenfit}), providing further validation of the model.

{\bf\textit{Soft excess}}\\
An intriguing characteristic of many HMXB systems is that their spectra show an excess of emission below 0.5 keV, known as the soft excess. Its origin is still not clear and might differ between sources, having several different contributions \citep{1993AAS...183.6502H,2004ApJ...614..881H}.

At first glance, the Absorbed and Orb 32 spectra exhibited an excess around 2 keV, initially interpreted as either a black body component or a reflection of the continuum. However, upon closer inspection, testing different data binning, a set of emission lines became discernible. One possibility then is that part of the soft excess is produced by unresolved emission lines generated in the stellar wind. This bump is overwhelmed by the NS emission (\texttt{bmc} component) in less absorbed spectra. 

In our case, we successfully fitted this excess emission with an absorbed low temperature \texttt{bbody} ($kT=0.06$ keV and $N^{22}_{\rm H}=0.8$). A straightforward candidate for this emission is the companion star. However, the luminosity expected for a B companion star should be $\sim\times 10^{32}$ erg s$^{-1}$ while the unabsorbed luminosity for this component turns out to be $\sim \times 10^{3}$ higher (see Table \ref{tab:model_kmeans} and \ref{tab:model_orbital}). 

Throughout the orbital spectra, the radius of this black body and its normalization are very similar, but in the \texttt{K-means} classification a progression can be observed, being enhanced during the Bright spectra and more suppressed in the Faint spectra. This is consistent with the scenario in which (at least part of) the soft excess are unresolved emission lines in the stellar wind, enhanced by a higher X-ray illumination from the compact object.

{\bf\textit{Fe K$\alpha$ line}}

The Fe\,K$\alpha$ emission line in 4U0114$+$65 is much weaker than normally found in HMXBs. The companion in 4U 0114$+$65 (B1Ia) is the coolest among the supergiant X-ray binaries \citep{2015A&A...579A.111K}. With a $T_{\rm eff}=24$ kK, 4U 0114$+$65 is in the middle of the bistability jump \citep[see low panel in Figure 3 presented by][]{1999A&A...350..181V}. This jump is caused by the sudden change in the ionization balance of Fe (\textsc{iii, iv}), which is a major contributor to the acceleration of the wind. \citet{2017A&A...606A.145S} suggest that this could have a strong impact on the efficiency of the mechanism to form and/or destroy clumps concluding that 4U 0114$+$65 have the typical thick wind of a supergiant star but with a much lower degree of clumping.

\begin{figure*}
\centering
\subfigure{\includegraphics[trim={13cm 8cm 5cm 0cm},width=0.9\textwidth]{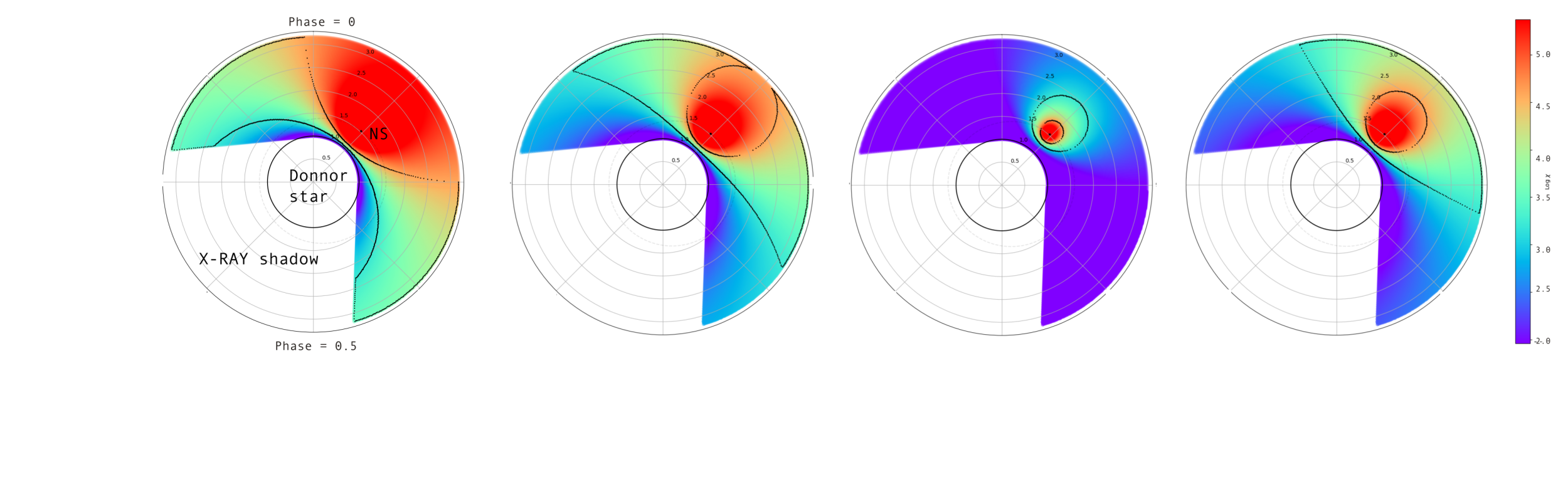}}
\caption{Ionization map in the plane of the orbit. From left to right of the Bright, Intermediate, Faint and Absorbed spectra respectively. The delimited regions represent the Fe K$\alpha$ region, i.e. with an ionization parameter lower than 100 and higher than 25. Please note that the orbital phase is ignored as the spectra is not phase-resolved. For visualization purposes, the orbital phase $\phi=0.12$ is used.} 
\label{fig:chimap}%
\end{figure*}

\begin{figure}

\subfigure{\includegraphics[trim={0cm 0cm 0cm 0cm},width=0.9\columnwidth]{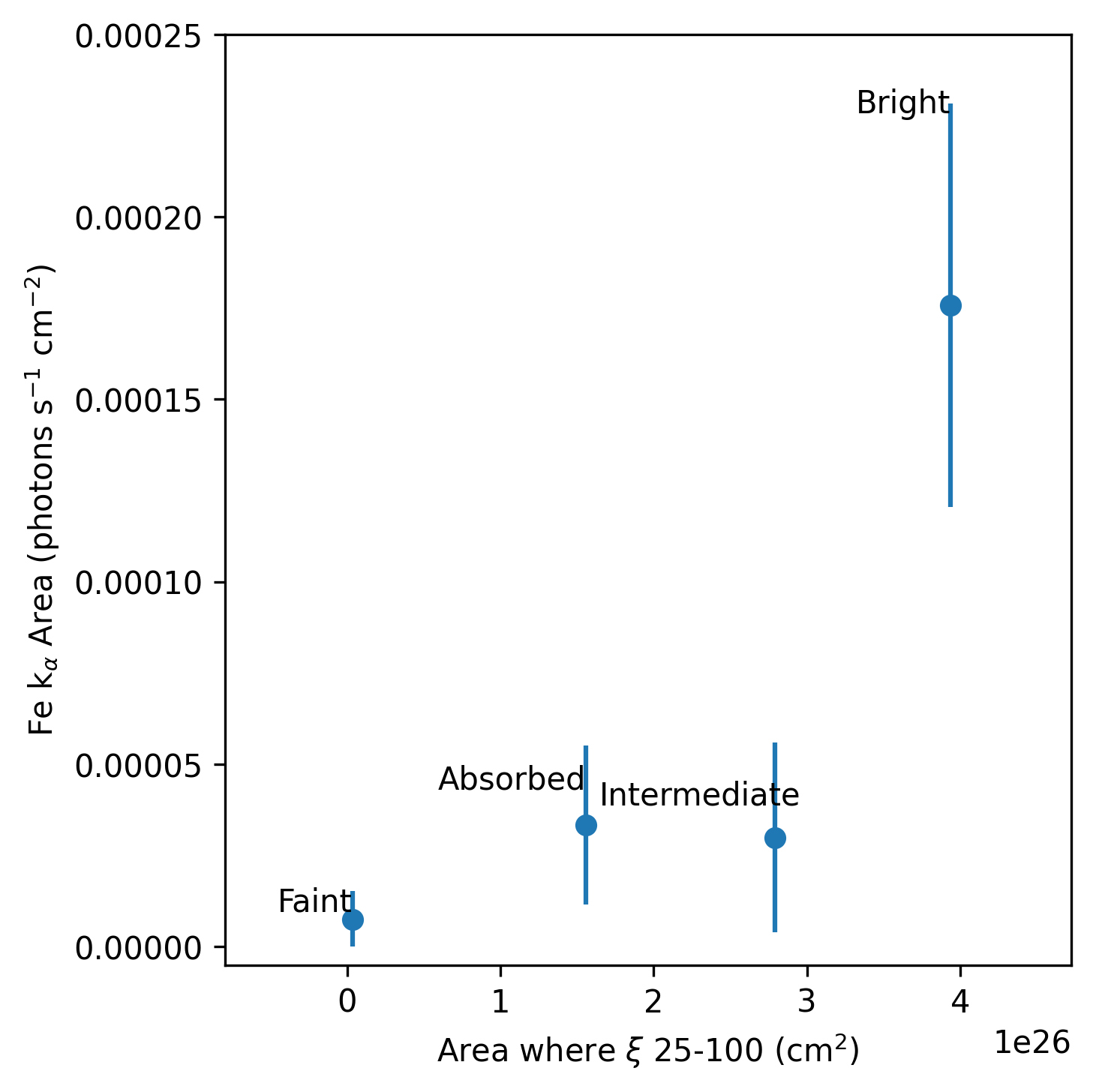}}
\caption{Fe K$\alpha$ formation region (ionization parameter in the range  $50-100$) vs the Fe K$\alpha$ line area.}
\label{fig:feka_radius}%
\end{figure}

 In order to have a strong Fe K$\alpha$ line, the ionization parameter $\xi=L_{\rm X}/n(r_{\rm X}) r_{\rm X}^{2}$, where $n(r_{\rm X})$ is the local particle density and $r_{\rm X}$ is the distance from the X-ray source (the NS), must be $\leq 10^{2}$. If we assume the former calculated mass-loss rate of $8.6 \pm 1.7 \times 10^{-7}$ $M_{\sun}$ yr$^{-1}$ the distance to the NS at which $\xi$ equals to 100 is 0.4, 0.3, 0.1 and 0.2 $R_{\star}$ for the Bright, Intermediate, Faint and Absorbed spectra respectively. This indicates that, within the Bright spectra, the X-ray source is powerful enough to illuminate the gas at distances of at least 0.4 $R_{\star}$ from the compact object.

An ionization map is shown in Fig. \ref{fig:chimap}, in which the Fe K$\alpha$ formation area (i.e. $\xi$ between 25 and 100) is represented by the delimited region inside the map. It is large in the Bright spectra and very compact in the Faint spectra. The area size display a positive trend with the Fe K$\alpha$ intensity (Fig. \ref{fig:feka_radius}). Please note that, in this particular case, the orbital phase is ignored because the spectra are classified using \texttt{k-means} clustering and are not phase-resolved. For visualization purposes, the orbital phase $\phi=0.12$ is used.

\newpage
{\bf\textit{NS pulse to pulse variability: a change of accretion regime.}}

The observed pulse-to-pulse luminosity variations and the spikes present within the light curve of 4U 0114$+$65 can be explained within the framework of the theory of quasi-spherical settling accretion onto slowly rotating magnetized NS \citep[see][]{2012MNRAS.420..216S} and its further development in subsequent works \citep{2014EPJWC..6402001S}. A comprehensive overview of this theory and its applications can be found in the review articles by \citet{2015ARep...59..645S} and \citet{2017arXiv170203393S}.

In sources with moderate X-ray luminosity (below $\sim 4 \times 10^{34}$ erg s$^{-1}$), a hot convective quasi-spherical shell forms above the NS's magnetosphere. The matter enters the magnetosphere due to Rayleigh-Taylor instability (RTI) at the magnetospheric boundary. This accretion processes in HMXRB has been known and studied for a long time (see \cite{1976ApJ...207..914A} and \cite{1977ApJ...215..897E}). The settling velocity in this regime is approximately two times smaller than the free-fall velocity, and the RTI is mediated by Compton and radiative plasma cooling.

We hypothesize that long spikes duration $t_B$, reflects the variability in the density and velocity of the gravitationally captured stellar wind and should scale with the free-fall time from the Bondi radius:

\begin{equation} t_B = \frac{R_B^{3/2}}{(GM)^{1/2}} \sim 2 \times 10^3 \mathrm{s}  \end{equation}

The spread in this timescale can be attributed to variations in the stellar wind velocity, which influences the Bondi radius and, consequently, the duration of $t_B$. For instance, a 10\% variation in the wind velocity results in approximately a 30\% variation in $t_B$, as the Bondi radius is inversely proportional to the square of the wind velocity.


A shorter timescale of variability, where $t_A$ represent short spike durations, is produced by the RTI near the magnetosphere. The growth rate of the RTI scales as:

\begin{equation} \sigma = \frac{1}{t_A} = \sqrt{Agk}, \end{equation}

where $A$ is the Atwood number (assumed to be close to 1 in our case), $g = \frac{GM}{R_A^2}$ is the gravitational acceleration at the magnetospheric boundary, and $k = 2\pi/\lambda$ is the RTI wave number. The most unstable modes scale with $R_A$, giving:

\begin{equation} t_A \sim \frac{R_A^{3/2}}{(GM)^{1/2}}. \end{equation}

Plugging our system's Alfven radius, a 1.4 $M_{\sun}$ NS mass and for $A \sim 1$, we get $t_A \sim 150$ s, in agreement with the observed spike duration (see Fig. \ref{fig:spikes_orbit_gm}). Since the magnetospheric radius is inversely proportional to the luminosity of the source, a negative correlation is expected between $t_A$ and the brightness, and this what is observed (Fig. \ref{Fig:duration_brightness}, upper panel), supporting RTI in the magnetosphere as the most likely origin of these short-duration spikes.

\begin{figure} \centering \includegraphics[trim={0cm 0cm 0cm 0cm},width=1\columnwidth]{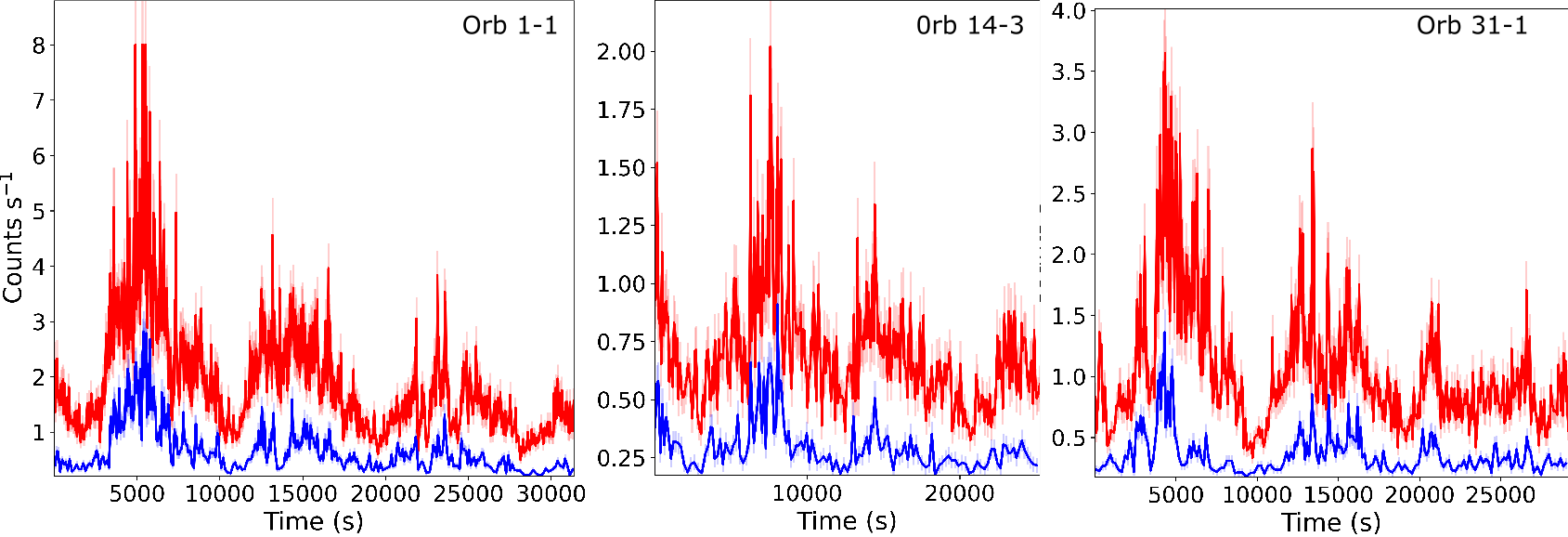} \caption{Light curves showing the decreasing pulse intensity trend (Orb 1-1, Orb 14-3, and Orb 31-1).} \label{fig:cycle.png} 
\end{figure}

The light curve exhibits a cyclic pattern where bright pulses are followed by intermediate and faint ones, with bright luminosity being up to 14 times higher than faint luminosity. Eventually, the intensity of the pulse recovers, and the cycle begins again (see Fig. \ref{fig:cycle.png}).

In this scenario, matter accumulates within the magnetospheric boundary until the pressure triggers more efficient Compton cooling and a faster RTI, leading to increased accreted matter flow and thus, the source brightness. As the settling velocity is approximately two times smaller than the free-fall velocity, the accumulated matter is gradually depleted, reducing the accretion rate and thus the luminosity until sufficient matter accumulates again, triggering another phase of the efficient Compton cooling. Indeed, considering the Atwood number \( A \sim 1 = \frac{\rho_{1} - \rho_{2}}{\rho_{1} + \rho_{2}} \), where \( \rho_{2} \) is the density inside the magnetosphere and \( \rho_{1} \) represents the density outside the magnetosphere and assuming \( \rho_{2} \) remains approximately constant with \( \rho_{2} \ll \rho_{1} \), an increase in \( \rho_{1} \) (corresponding to greater matter accumulation above the Alfven radius) would result in a shorter \( t_A \), and vice versa. This behavior aligns with our observations (shorter spikes in brighter spectra, corresponding with high accumulation of matter and longer spikes in faint spectra, when the matter above the magnetosphere is depleted).

The spectral parameters derived form the spectral model support this scenario. The local absorption column $N_{\rm H}$ is $\sim 2$ times larger in the Bright spectra than in the Faint spectra, suggesting an accumulation of circumstellar matter (see Table \ref{tab:model_kmeans}). The $\alpha$ parameter from the \texttt{bmc} component changes significantly, from $\sim 0.1$ in the case of the Bright spectra, indicating a very efficient Comptonisation mechanism compared to the Faint spectra $\sim 4$.

\section{Summary and conclusions.}

In this analysis of the 4U 0114+65 system, based on 200 ks of \textit{Chandra} observations, we examine the variability in the accretion process of one of the slowest pulsars known. The key results are:

\begin{itemize}

    \item \textit{Circumstellar absorption.}  The system's inclination angle with respect to the observer and the companion's mass-loss rate were successfully constrained by fitting the variation of N$_H$ through the orbit using a PSO algorithm. The result suggests an inclination of $40\pm 3$ $^{\circ}$ and a lower limit for the mass-loss rate of $8.6 \pm 1.6 \times 10^{-7}$ $M_{\sun}$ yr$^{-1}$, in alignment with previous studies.\\
    
    \item \textit{Soft excess}. The soft excess present in the X-ray spectrum, especially around 2 keV, is at least, partly formed by unresolved emission lines. This feature could only be revealed in the more absorbed spectra, when the emission produced by the NS at low energy ranges is partly masked by the circumstellar absorption.\\
    
    \item \textit{Fe K$\alpha$ line emission site.} This emission line, generally strong within HMXRBs, is unusually weak in this source. We hypothesize that 4U 0114+65 has the typical thick wind of a supergiant star but with a much lower degree of clumping. In order to have a strong Fe K$\alpha$ line, the ionization parameter $\xi$ must be $\leq 10^{2}$. This ionization parameter is reached at distances from the NS of 0.4, 0.3, 0.1 and 0.2 $R_{\star}$ for the Bright, Intermediate, Faint and Absorbed spectra respectively. This indicates that, within the Bright spectra, the X-ray source is powerful enough to illuminate the gas at distances of at least 0.3 $R_{\star}$ resulting in a higher Fe K$\alpha$ line flux.\\ 
    
    \item \textit{Accretion regime cycle.} The pulse-to-pulse variability and spikes in the light curve can be explained within the theory of quasi-spherical settling accretion. Longer-duration spikes correspond to variations close to the Bondi radius, while shorter spikes are likely caused by RTI near the magnetosphere. As the settling velocity is approximately two times lower than the free-fall velocity, matter accumulates within the magnetospheric boundary until the pressure triggers efficient Compton cooling, leading to increased source brightness. The accumulated matter is gradually depleted, reducing the accretion rate and thus the luminosity until sufficient matter accumulates again, triggering another phase of efficient Compton cooling.\\
    \end{itemize}

\section*{Data Availability}
The data analyzed in this study can be found in the {\it Chandra} archive under the observation identification numbers 23432, 24479, 24480, 24481, 24482, 24483, 16177, 26178  and 16180.  Appendices can be found in zenodo: \href{}{https://zenodo.org/records/14645709}.

\section*{Acknowledgements}

This research has been funded by the ASFAE/2022/02 project from the Generalitat Valenciana. N. Schulz and M. Nowak were supported by NASA Chandra grants GO3-24018A and GO3-24018B, respectively. We acknowledge the constructive criticism of the referee whose comments improved the content of the paper.



\bibliographystyle{aa}
\bibliography{example.bib} 
\newpage

\onecolumn




\label{lastpage}
\end{document}